\documentclass{aa}  
\usepackage{graphicx}
\def\kms    {\ifmmode{{\rm ~km\,s}^{-1}}\else{~km\,s$^{-1}$}\fi}
\def\arcm   {$^{\prime}$}
\def\arcs   {$^{\prime\prime}$}
\usepackage{txfonts}
\usepackage[]{natbib}
\bibpunct{(}{)}{,}{a}{}{,} 
\begin{document}
   \title{The AMIGA sample of isolated galaxies}

   \subtitle{IV. A catalogue of neighbours around isolated galaxies\thanks{Full Table~\ref{tab:comp} is available in electronic form at the CDS via anonymous ftp to {\tt cdsarc.u-strasbg.fr (130.79.128.5)} or via {\tt http://cdsweb.u-strasbg.fr/cgi-bin/qcat?J/A+A/vvv/ppp} and from {\tt http://www.iaa.es/AMIGA.html}.}}

\titlerunning{The AMIGA sample of isolated galaxies. IV.}

   \author{S. Verley\inst{1,2,3}
	\and
	S. C. Odewahn
	\inst{4}
	\and
	L. Verdes-Montenegro
	\inst{2}
	\and
	S. Leon
	\inst{5,2}
	\and
	F. Combes
	\inst{1}
	\and
	J. Sulentic
	\inst{6}
	\and
	G. Bergond
	\inst{2}
	\and
	D. Espada
	\inst{2}
	\and
	E. Garc\'ia
	\inst{2}
	\and
	U. Lisenfeld
	\inst{7}
	\and
	J. Sabater
	\inst{2}
          }


   \institute{LERMA -- Observatoire de Paris, 61, avenue de l'Observatoire, 75014 Paris, France\\
              \email{Simon.Verley, Francoise.Combes@obspm.fr}
        \and
             Instituto de Astrof\'isica de Andaluc\'ia -- CSIC, C/ Camino Bajo de Huetor, 50, 18008 Granada, Spain\\
             \email{simon, lourdes, jsm, gilles, daniel, garcia@iaa.es}
	\and
	     INAF -- Osservatorio Astrofisico di Arcetri, Largo E. Fermi, 5 I-50125 Firenze, Italy\\
	     \email{simon@arcetri.astro.it}
	\and
	     McDonald Observatory, University of Texas, USA\\
	     \email{sco@astro.as.utexas.edu}
	\and
	     Instituto de RadioAstronom\'ia Milim\'etrica, Avenida Divina Pastora, 7, N\'ucleo Central, E 18012 Granada, Spain\\
	     \email{leon@iram.es}
	\and
	     Department of Astronomy, University of Alabama, Tuscaloosa, USA\\
	     \email{giacomo@merlot.astr.ua.edu}
	\and
	     Departamento de F\'\i sica Te\'orica y del Cosmos, Facultad de Ciencias, Universidad de Granada, Spain\\
	     \email{ute@ugr.es}
             }

   \date{Received; accepted}

  \abstract
   {Studies of the effects of environment on galaxy properties and evolution require well defined control samples. Such isolated galaxy samples have up to now been small or poorly defined. The AMIGA project ({\bf A}nalysis of the interstellar {\bf M}edium of {\bf I}solated {\bf GA}laxies) represents an attempt to define a statistically useful sample of the most isolated galaxies in the local (z$\leq$0.05) Universe.
}
   {A suitable large sample for the AMIGA project already exists, the Catalogue of Isolated Galaxies (CIG, Karachentseva 1973; 1050 galaxies), and we use this sample as a starting point to refine and perform a better quantification of its isolation properties.
}
   {Digitised POSS-I E images were analysed out to a minimum projected radius R~$\geq$~0.5~Mpc around 950 CIG galaxies (those within V$_r$~=~1500~\kms\ were excluded). We identified all galaxy candidates in each field brighter than B~=~17.5 with a high degree of confidence using the LMORPHO software. We generated a catalogue of approximately 54\,000 potential neighbours (redshifts exist for $\approx 30$\% of this sample).
}
   {Six hundred sixty-six galaxies pass and two hundred eighty-four fail the original CIG isolation criterion. The available redshift data confirm that our catalogue involves a largely background population rather than physically associated neighbours. We find that the exclusion of neighbours within a factor of four in size around each CIG galaxy, employed in the original isolation criterion, corresponds to $\Delta$V$_r \approx 18000$~\kms\ indicating that it was a conservative limit.
}
   {Galaxies in the CIG have been found to show different degrees of isolation. We conclude that a quantitative measure of this is mandatory. It will be the subject of future work based on the catalogue of neighbours obtained here.
}


   \maketitle
%
\section{Introduction}

During the past 30 years it has become clear that galaxy properties (e.g. morphology, star formation, nuclear activity) and evolution may be driven as strongly by environment as by initial conditions. The role of environmental conditions is not yet fully quantified for at least two reasons: 1) confusion about the definition of ``environment'' and 2) lack of control samples of galaxies minimally affected by environment. The former confusion arises because there are two kinds of (observable) environmental influences: a) one-on-one and b) local galaxy surface density. A single, sometimes difficult to identify, neighbour can be capable of a larger effect than an enhanced local galaxy surface density. Effects related to local galaxy surface density can be especially difficult to quantify because automated sample selection can often miss these close neighbours. Ideally we seek a statistically useful sample of galaxies that has been carefully cleaned of: a) close neighbours and that reside in b) the lowest galaxy surface density regions of the local Universe. In order to be statistically useful the sample must be large enough to allow us to assess environmental effects both as a function of morphological type and luminosity. The motivation of the AMIGA ({\bf A}nalysis of the Interstellar {\bf M}edium of {\bf I}solated {\bf GA}laxies\footnote{\tt http://www.iaa.es/AMIGA.html}) project is to identify such a sample of isolated galaxies.

The AMIGA project adopted the Catalogue of Isolated Galaxies \citep[CIG:][]{1973AISAO...8....3K} as a starting point. The strength of the CIG involves its size (1050 galaxies) and its selection with a strong isolation criterion. Redshifts are available for almost the entire sample, which is large enough to allow severe refinement without reducing the resulting catalogue to a few tens of galaxies. Previous papers in this series included: 1) improvement in positional accuracy \citep{2003A&A...411..391L}, 2) optical characterisation including derivation of the optical luminosity function \citep[AMIGA-I:][]{2005A&A...436..443V}, 3) morphological revision using POSS-II (and SDSS overlap) and type-specific OLF analysis \citep[AMIGA-II:][]{2006A&A...449..937S} and 4) mid- and far-infrared properties using the IRAS database \citep[AMIGA-III:][]{2007A&A...462..507L}. Studies of the radio continuum, HI \citep{2005A&A...442..455E,2006.PhD.Thesis.E}, CO and H$\alpha$ emission \citep{2005.PhD.Thesis.V} properties are in progress along with a study of the small AGN population found in the sample. This paper focuses on a reassessment of the isolation degree for all galaxies in the CIG with V$_r \geq 1500$~\kms. A different assessment strategy is required for the nearest galaxies in our sample which are all part of the local supercluster.

In our previous works we identified several CIG galaxies failing \citeauthor{1973AISAO...8....3K}'s criterion, hence motivating us to carefully revise the isolation of the CIG members. Here we perform a census of the environment of the most isolated galaxies in the local Universe (within $\sim$350~Mpc) located in the northern hemisphere. In Sect.~\ref{sec:prev}, we review previous work on isolated galaxies. In Sect.~\ref{sec:CIG}, we present in detail the CIG as well as several revisions and improvements performed in the bibliography. We also illustrate the isolation definition using the Milky Way as an example. In Sect.~\ref{sec:pipe}, we describe in detail the method used to revise the isolation of the CIG galaxies, including a description of our automated pipeline used to produce a catalogue of their potential neighbours. We have also compiled redshifts for these possible neighbours from available databases, as we explain in Sect.~\ref{sec:red}. In Sect.~\ref{sec:disc}, we revise \citeauthor{1973AISAO...8....3K}'s catalogue in order to determine how many galaxies still remain isolated based on our new data. We present our conclusions in Sect.~\ref{sec:conc}. From our study we conclude that a quantification of the isolation is needed; this will be presented in a future article \citep[in prep.]{2007A&A...2..2V}.



\section{Previous work on samples of isolated galaxies} \label{sec:prev}

Interest in isolated galaxy compilations increased in the 1970s-80s as evidence accumulated that mergers, interactions or simply high local galaxy environmental density can play an important role in observed galaxy properties and evolution \citep{1972ApJ...178..623T,1976ApJS...32..171S,1977ARA&A..15..437T,1978ApJ...219...46L,1978AJ.....83..322S}. As recently as 1975 the consensus was against an interaction induced signature in interacting galaxies \citep[e.g.,][]{1973Natur.241..260A}. The CIG was criticised as a poor field sample because it obviously lacked global homogeneity \citep{1983ApJ...275..472H}. Given the identifiable components of the CIG mentioned above it is not surprising that the full CIG sample failed a covariance analysis \citep{1986A&A...154..343V}. The latter study offered an alternative, albeit small (43 galaxies), catalogue of very isolated galaxies. Paradoxically only one CIG galaxy (CIG~319) was included in this automated compilation while many were found to be components of isolated binary galaxies (Catalogue of Pairs of Galaxies, CPG) compiled by \citet{1972SoSAO...7....1K} in a companion survey to the CIG. This result illustrates the danger inherent in compiling 2D (or 3D) catalogues of isolated galaxies with a sharp magnitude cutoff. Such catalogues will often include first ranked galaxies in cluster cores as well as close (especially hierarchical) binaries where one component falls just below the cutoff magnitude of the catalogue employed. The one clear result to emerge from the above effort is that CIG~319 is likely to be a very isolated galaxy.

A debate on the nature of the spatial distribution of galaxies took place in the mid-1970s: using the covariance function of the distribution of galaxies, \citet{1974ApJ...189L..51P, 1974A&A....32..197P} found no evidence of an initially homogeneous component of the galaxy population and, on the contrary, endorsed the view of hierarchical series of densities. However, studying galaxies brighter than 14$^{\rm th}$ magnitude, \citet{1975ApJ...197L..89T} found two distinct populations, one strongly clustered and a population of ``single'' galaxies (32\%) distributed homogeneously on scales $\leq 20$ Mpc. But \citet{1977ApJ...211....1S} showed that the previous sample did not constitute a true field population and if such a population existed, it amounted to substantially less than 18\% in a catalogue selected by apparent magnitude. \citet{1977ApJ...216..694H} revised the \citeauthor{1975ApJ...197L..89T} sample down to a fainter magnitude (15.7 mag) and found that isolated galaxies could only represent 3.6\% of all the galaxies. \citet{1986A&A...154..343V} also emphasised that isolated galaxies did not exist in an absolute sense because clustering on large scale dominates in all regions of space (for small redshift at least).

Studies comparing redshifts of isolated galaxies with redshifts of groups confirmed that isolated galaxies generally belong to groups, but at such large distances from their centres ($\sim 4$ Mpc) that they have not undergone any physical influence from these groups \citep{1981A&A....97..223B}. \citet{1983ApJ...275..472H} showed that likewise most of the isolated galaxies are outer components of groups or clusters.

Hence, it seems difficult to find a truly isolated population of galaxies, but instead one can have access to regions of very low galaxy density, where the galaxies reflect properties characterising their formation. However, during the past 30 years, a variety of widely different criteria has been used when defining isolation (magnitude limited samples, redshift information used or not, distance to the nearest galaxies different from one definition to the other, etc.), as shown by the abundant literature: \citet{1975ApJ...197L..89T,1981A&A....97..223B,1986A&A...154..343V,1993ApJ...405..464Z,2001AJ....122.2923A,2001AJ....121..808C,2002ApJS..142..161P,2003ApJ...598..260P,1996A&AS..120....1M,1999A&A...344..421M,2002A&A...393..389M,2003Ap&SS.284..711M,2004A&A...420..873V}. Most of these studies only sample ten to approximately two hundred galaxies, which is not sufficient for statistical analysis.

Studies with independent isolated samples usually involve small numbers and show a surprisingly small overlap with the CIG \citep[e.g.,][]{1991AJ....102.1980X,1996A&AS..120....1M,1999A&A...344..421M,1998MNRAS.295...99M,1999A&A...351...43A,1999AJ....117.2168P,2001AJ....121..808C,2001AJ....121.1358K,2002ApJS..142..161P,2004ApJ...607..810M,2004MNRAS.354..851R}.
In a few cases the samples include southern objects that lie outside the CIG sky coverage. Many of them, surprisingly, contain more overlap with catalogues of pairs or triplets than with the CIG. This usually involves computer-based compilations from a magnitude limited (2D or 3D) catalogue \citep{1975ApJ...197L..89T,1986A&A...154..343V}. If one of the galaxies in a pair falls below the magnitude limit of the catalogue then the pair will be adopted as an isolated single galaxy more readily than a CIG member. The pair isolation criterion is more stringent if isolation is defined in terms of pair separation rather than a component galaxy diameter. Visual confirmation of isolation is essential and is one of the strengths of the CIG. This list is not intended to argue about the relative merits of different isolated galaxy selection criteria. It is intended to show: 1) the lack of consensus about what constitutes a reasonable isolated galaxy or isolated galaxy sample, 2) how difficult it can be to compare different results for different selected isolated samples and 3) the confusion about selection on the basis of nearest neighbour vs. selection on the basis of local surface density.

Most galaxies in other samples that are found north of $\delta = -3^\circ$, and are missing from the CIG, reflect violation of the CIG isolation criterion rather than having been overlooked during the CIG compilation. The CIG isolation criterion is more stringent than most others that have been used. One of the goals of the AMIGA project is to extract from the CIG a significant subsample of the most isolated galaxies which must represent the low density tail of galaxy population in the local Universe. The remainder of the sample will involve degrees of lesser isolation where effects of environment might begin to be detected. One of our goals is to detect that threshold.

\section{The Catalogue of Isolated Galaxies} \label{sec:CIG}

\subsection{Definition}

The catalogue is composed of 1051 objects with apparent Zwicky magnitude $m_{\rm zw}$ brighter than 15.7 and declination $> -3^{\circ}$. \citeauthor{1973AISAO...8....3K} visually inspected the Palomar Sky Survey prints, trying to identify those galaxies in the Catalogue of Galaxies and Clusters of Galaxies \citep[CGCG,][]{1968cgcg.book.....Z} which have no near similar size neighbours. Primary galaxies with angular diameter $D_p$ are considered \textit{isolated} if any neighbour with diameters $D_i$ (with $D_p/4 \le D_i \le 4D_p$) has an apparent angular separation $R_{ip}$, from the primary galaxy, greater than $20D_i$.



This criterion statistically implies that all possible effects of a past interaction on the morphological or dynamical properties of a CIG galaxy, or those concerning the enhancement of star formation processes, have likely been erased at the present time. Because this represents a lower limit on the time since the last interaction between a CIG galaxy and a potential neighbour, the CIG galaxies have apparently been isolated for much (if not all) of their existence. For instance, for a CIG galaxy with $D_p = 3'$, no neighbour with $D_i = 12'$ may lie within 240\arcm\ and no neighbour with $D_i = 0\farcm75$ may lie within 15\arcm. If one assumes an average $D_p$ = 25 kpc for a CIG galaxy and a typical ``field'' velocity V = 150 km s$^{-1}$ then an approximately equal mass perturber would require 3$\times$10$^9$ years to traverse a distance of $20 D_i$ \citep{1978AJ.....83..348S}. 

This is a conservative criterion in the sense that, since no redshift data is used for the isolation definition, a truly isolated galaxy may be excluded from the CIG due to a projected background/foreground neighbour: galaxies isolated in space do not necessarily appear isolated in the sky. As a result of these projection effects the CIG is not fully complete. Nevertheless, the sample is still reasonably complete, according to the \citet{1968ApJ...151..393S} luminosity volume test which gives $<$$V/V_m$$>$ = 0.42 at a Zwicky magnitude of 15.0 \citep{1977ApJ...216..694H,1991ApJ...374..407X,1999AJ....118..108T,2005A&A...436..443V}. On the other hand all galaxies that are included should be isolated. The CIG is a sample of galaxies isolated from similarly sized neighbours, but it is clear that dwarf neighbours are not excluded.

Several refinements of the CIG have been performed since its selection. \citet{1980SvA....24..665K} discussed her isolation criterion and found that 24 galaxies (with known radial velocities) passed the isolation criterion and belong to pairs, groups, or clusters. Other authors \citep{1978AJ.....83..348S, 1984AJ.....89..758H, 1991ApJ...374..407X} reported that some CIG galaxies are, in fact, members of interacting systems: CIGs 6, 7, 80, 197, 247, 278, 324, 347, 349, 444, 469, 559, 663, 781, 802, 809, 819, 850, 851, 853, 938, 940, 946, 1027, 1028.

\citet{1980AJ.....85.1010A} and Karachentseva (1986\footnote{Unpublished documentation supplied with the catalogue by the Centre de Donn\'ees Astronomiques, Strasbourg.}) refined the original isolation criterion by assigning the following codes:
\begin{itemize}
\item Code 0: Isolated according to \citeauthor{1973AISAO...8....3K} (902 galaxies);
\item Code 1: Marginally isolated (85 galaxies);
\item Code 2: Member of a group or cluster (64 galaxies).
\end{itemize}

A few detailed studies of CIG galaxies (recognised as very isolated) also exist, see for instance: CIG 947 \citep{1995A&A...300...65V}; CIG 121 \citep{1996A&A...310..722K}; CIG 710 \citep{1997A&A...321..754V}; CIGs 164, 412, 425, 557, 684, 792, 824, 870, 877 \citep{2004AJ....127.3213M}; CIG 96 \citep{2005A&A...442..455E}.

The CIG is complemented by catalogues of isolated pairs, triplets and compact groups (largely quartets); none of them take into account more hierarchical systems.


\begin{figure}
\includegraphics[width=\columnwidth]{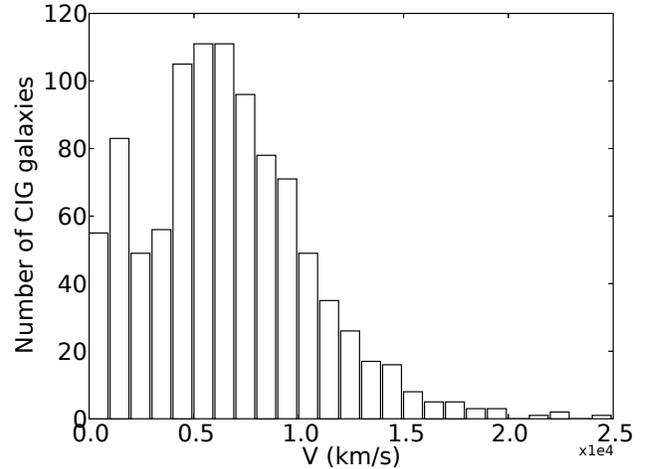}
\caption{Distribution of the recession velocities of the CIG galaxies.} \label{fig:histVelCig}
\end{figure}

\subsection{Would the Galaxy belong to the CIG?}

In order to illustrate \citeauthor{1973AISAO...8....3K}'s isolation criterion we have applied it to the Milky Way for which the distances and sizes of its neighbours are relatively well known \citep{2006astro.ph..5564G}.

The Milky Way is a common spiral galaxy (its stellar mass is about 10$^{11}$ M$_\odot$), with a disk of about 30~kpc in diameter. Hence, all the galaxies which would possibly violate the \citeauthor{1973AISAO...8....3K}'s criterion would have diameters between $\sim$ 7.5 kpc for the smallest and 120 kpc for the largest. As the neighbour galaxies can lie at a distance as large as 20 times their diameter away, we would have to check for all the members within 2.4 Mpc. Among the nearby groups of galaxies, only the Sculptor group (1.8 Mpc away) lies inside this limit, the others are all further than 3 Mpc, hence not concerned (M\,81: 3.1 Mpc; Centaurus: 3.5 Mpc; M\,101: 7.7 Mpc; M\,66 + M\,96: 9.4 Mpc; NGC\,1023: 9.5 Mpc; etc.). The Sculptor group has six members: NGC\,253 (diameter of 14.4 kpc), the brightest galaxy of the group, would not violate the isolation criterion.

Hence, the question of the isolation of the Milky Way would only involve galaxies of the Local Group. Our galaxy's brightest satellite systems are the Magellanic Clouds. The Large Magellanic Cloud is 49 kpc away and has a diameter of $\sim$ 9.3 kpc: this neighbour violates \citeauthor{1973AISAO...8....3K}'s criterion. The Small Magellanic Cloud has a diameter of about 5.4 kpc and would not be taken into account by \citeauthor{1973AISAO...8....3K}'s criterion: we see here a limitation of the criterion which does not take into account dwarf companions, already mentioned in the previous section.

Belonging to the local Group but farther away, the Andromeda Galaxy (M\,31) has an apparent angular major diameter of 190\arcm, corresponding to about 40 kpc. Its influence would affect any galaxy as far as 800 kpc from it, according to the \citeauthor{1973AISAO...8....3K}'s criterion. Since the distance separating the Milky Way from the Andromeda galaxy is about 725 kpc, this latter would also violate \citeauthor{1973AISAO...8....3K}'s criterion. On the other hand, the Triangulum galaxy (M\,33) is about 840 kpc away and, due to its relatively small diameter ($\sim$ 16.2 kpc), would not exert any noticeable influence on the Milky Way. This would be true if the system Milky Way-M\,33 would have been seen in the best case (the line of sight perpendicular to the plane defined by the two galaxies). If the system is seen from other points of view, the apparent distance separating the two galaxies will become smaller and reach a point where the Milky Way would no longer appear isolated relative to M\,33. This illustrates the above referred effect of incompleteness induced by a strong definition of isolation, depending on apparent 2D distances.

\section{The AMIGA revision} \label{sec:pipe}

Despite the various revisions by the authors cited in Sect.~\ref{sec:prev}, we chose to improve \citeauthor{1973AISAO...8....3K}'s sample by checking in an automated, homogeneous way the isolation of the galaxies and by listing/classifying the neighbour galaxies.

\subsection{The sample}

We have excluded from our revision all the CIG galaxies with radial velocities lower than 1500\kms\ (100 galaxies, see Fig.~\ref{fig:histVelCig} for the velocity distribution of all the CIG galaxies) since, as pointed out by \citet{1978AJ.....83..348S} and \citet{1984AJ.....89..758H}, the area searched for potential neighbours of nearby CIG galaxies is spread over a large surface on the sky, which makes the search overwhelming. Our final target sample is composed of 950 CIG galaxies.

\subsection{Data analysis}

We developed an original method to check the isolation of the CIG galaxies. This work was motivated by the fact that objects brighter than an apparent magnitude $m_B \approx 17.5$ are mis-classified at a high rate in present on-line reductions of the all-sky Schmidt surveys. Fainter than $m_B \approx 17.5$, the mean isophotal surface brightness of stars begins to be comparable with that of many galaxies, and the number of pixels per source at this level (assuming a typical isophotal threshold of $\mu_B \approx 23.5$) is too small to unambiguously differentiate stars and galaxies on the basis of shape. In the following, we describe the method used to reliably identify bright (i.e. $m_B < 17.5$) galaxies around our CIG fields of interest.

\subsubsection{Size of the studied fields}

In order to recover the bright galaxies with high success rate, we reduced bright image classification in our CIG fields using Palomar Observatory Sky Survey (POSS-I E, central wavelength = 6510 \AA) images obtained with the Digitised Sky Survey (DSS). We have assembled a software pipeline for producing star/galaxy catalogues in the area around each CIG field. The digital images have a pixel size of 25 $\mu$m (1\farcs7/pixel).

\begin{figure}
\includegraphics[width=\columnwidth]{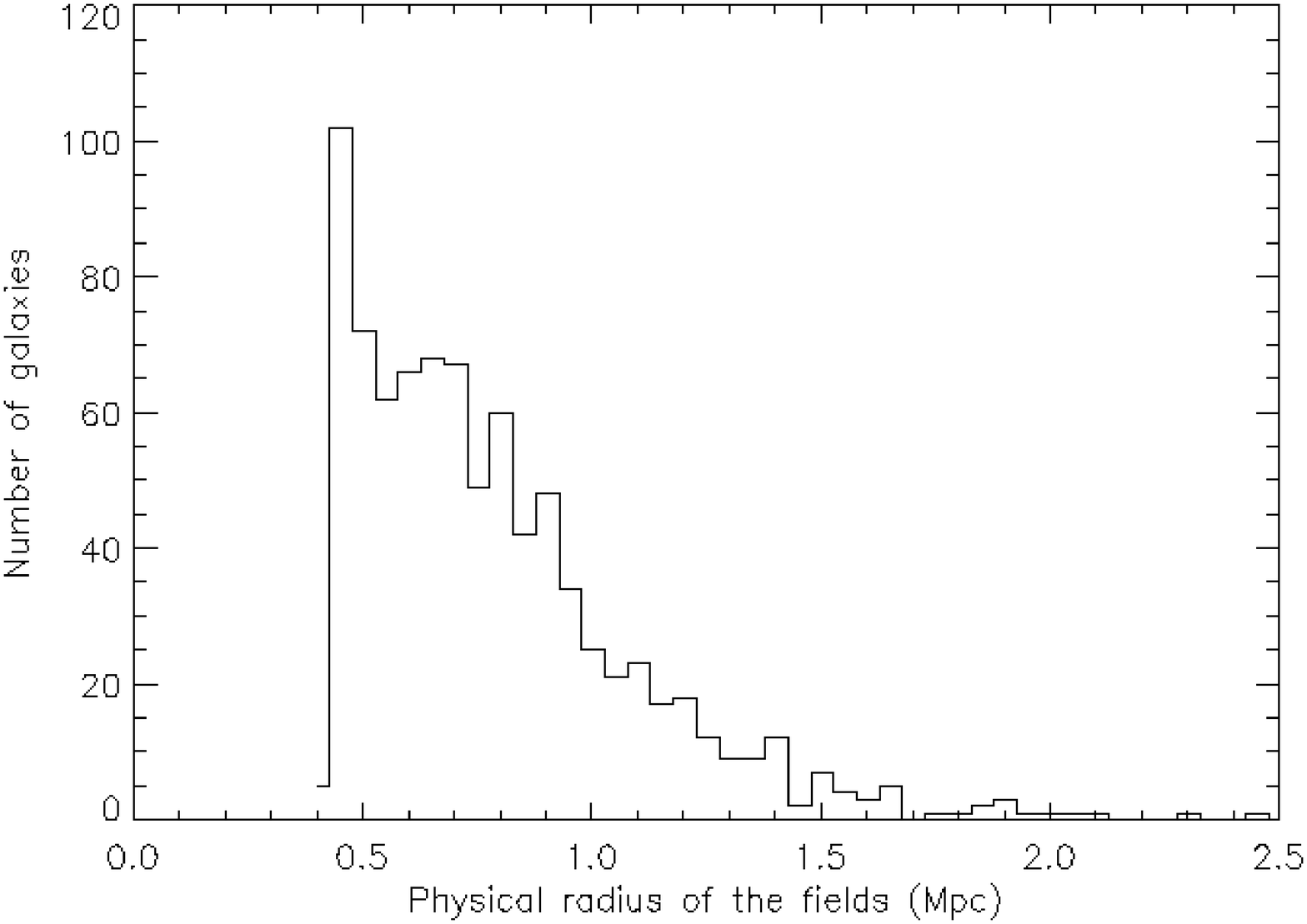} \caption{Physical radius of the fields inspected for our CIG sample (the velocity is available for 888 CIG galaxies).} \label{fig:tamFis856}
\end{figure}

We chose to evaluate the isolation degree in a minimum physical radius of 0.5 Mpc ($H_0$ = 75 \kms\ Mpc$^{-1}$), centred on each CIG galaxy (see Fig.~\ref{fig:tamFis856}). Assuming a field velocity of 150\kms, it would require at least $3.2\times10^9$ years for a neighbour to travel over this radius. Due to pipeline capacity and server limits, we could not handle fields larger than $55' \times 55'$. To reach the physical radius of 0.5 Mpc, the fields requiring a larger size were composed of various $55' \times 55'$ fields, with a small strip overlapping between two adjacent fields. We developed a tool to keep only one source when an object was detected more than once in adjacent fields. Below, we show the number of CIG galaxies in each field size employed:
\begin{itemize}
\item 767 galaxies with $55' \times 55'$ ;
\item 134 galaxies with multi-fields $110' \times 110'$;
\item 49 galaxies with multi-fields $165' \times 165'$.
\end{itemize}

The $55' \times 55'$ fields concerned galaxies with an observed recession velocity greater than 4687\kms\ (including the 62 galaxies with no velocity data, see Table~\ref{tab:62noZ}); the $110' \times 110'$ multi-fields correspond to galaxies between 2343 and 4687\kms; the $165' \times 165'$ multi-fields to recession velocities between 1500 and 2343\kms.

\begin{table}
\begin{center}
\caption{List of the 62 CIG galaxies with unknown redshift.} \label{tab:62noZ}
\begin{tabular}{c c c c c c c}
\hline \hline
CIG & CIG & CIG & CIG & CIG & CIG & CIG \\
\hline
0003 & 0272 & 0479 & 0629 & 0717 & 0814 & 0899 \\
0017 & 0297 & 0535 & 0632 & 0729 & 0821 & 0908 \\
0026 & 0311 & 0558 & 0664 & 0730 & 0822 & 0964 \\
0035 & 0320 & 0583 & 0673 & 0737 & 0842 & 0968 \\
0046 & 0360 & 0587 & 0681 & 0765 & 0846 & 0977 \\
0048 & 0369 & 0594 & 0687 & 0774 & 0869 & 0995 \\
0070 & 0394 & 0597 & 0704 & 0787 & 0878 & 0996 \\
0254 & 0414 & 0607 & 0707 & 0790 & 0885 & 1049 \\
0263 & 0459 & 0628 & 0713 & 0804 & 0887 &      \\
\hline
\end{tabular}
\end{center}
\end{table}

\subsubsection{Detection of the sources}

We used SExtractor \citep{1996A&AS..117..393B} to detect the sources in the images, with a threshold 3 times higher than the root mean square of the background estimation. Before the source extraction we applied a Gaussian convolution with a full width at half maximum of 2 pixels and a size of $5 \times 5$ square pixels. Then, all the objects larger than 4 pixels were detected, which corresponds to a diameter smaller than 2 kpc at the typical distance of the CIG galaxies (corresponding to a velocity of about 6500\kms, see Fig.~\ref{fig:histVelCig}). 

\subsubsection{Star/galaxy separation}

\begin{figure}
\includegraphics[width=\columnwidth]{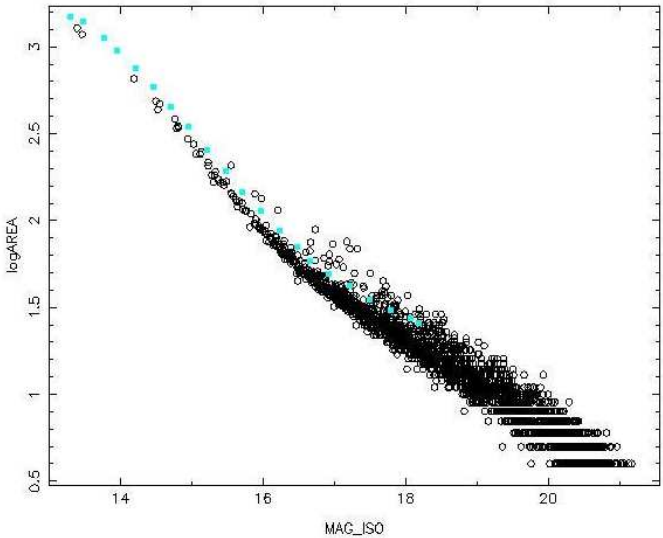}
\caption{Star/galaxy separation parameter plane. The objects above the separation line (blue points) are classified as {\sc Galaxy}, while the ones below are classified as {\sc Star}. The objects fainter than the extent of the separation line are not classified, their type is {\sc Unknown}.}\label{fig:paramSpace714}
\end{figure}

The images were reduced using AIMTOOL in LMORPHO \citep{1995PASP..107..770O, 1996ApJ...472L..13O, 2002ApJ...568..539O}, and a Graphical User Interface (GUI) driven star/galaxy separation procedure was used to classify detected sources as: {\sc Star}, {\sc Galaxy} or {\sc Unknown} (for the faint, low resolution sources). Star/galaxy separation was performed in the log(area) vs. magnitude (SExtractor MAG\_ISO), which was found to robustly isolate the stellar locus brighter than $M_B \approx 17.5$ in a random sample of Schmidt plates. A typical star/galaxy separation parameter plane from a POSS-I E image (CIG 714) is shown in Fig.~\ref{fig:paramSpace714}. The galaxies have a lower surface brightness than the stars and in the Log(area) vs. magnitude plane, the two classes of objects fall in different loci \citep{2000A&A...359..907L}. The stellar locus in Log(area) vs. magnitude plane was manually located using an interactive GUI approach because the shape and location of this locus changes significantly on different POSS-I Schmidt plates. All the points that lie above the curve defined by the blue filled circles (which is described with a cubic spline) were classified as {\sc Galaxy}. The points below this curve were classified as {\sc Star}. Points lying outside the spline range (brighter or fainter in magnitude than the extent of the red points) were classified as {\sc Unknown}. As a final step, we archived our catalogues in the form of ASCII files.

\subsubsection{Visual checks} \label{sssec:checks}

\begin{figure}
\includegraphics[width=\columnwidth]{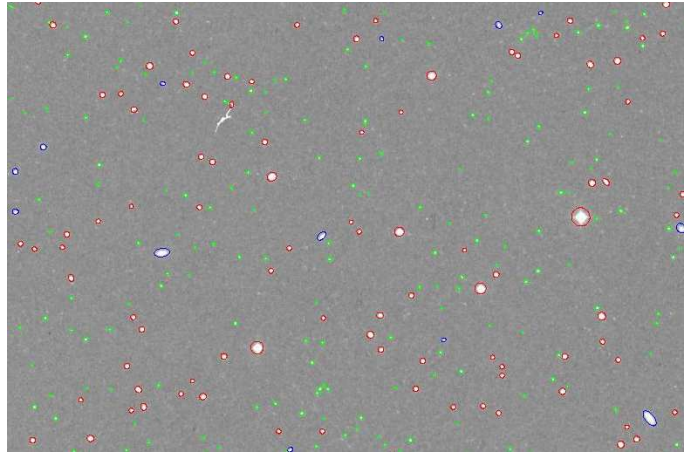}
\caption{Close-up view of the distribution of galaxies around CIG 714 (the bottom-right galaxy). The detected galaxies are marked with blue ellipses, the stars are marked by red ellipses. The objects too small or to faint to be assigned a type are shown by the green ellipses. The plate defects were removed from our object extraction.} \label{fig:field714}
\end{figure}

For a visual check, the GUI allows the user to view the image catalogue in the form of coloured-ellipse markers over-plotted on the DSS image (see Fig.~\ref{fig:field714}). The blue ellipses indicate the {\sc Galaxies} detected, the red ones over-plot the {\sc Stars} and the green circles mark the sources that were not classified. One of us (S. V.) systematically verified all the objects ({\sc Galaxy}, {\sc Star} and {\sc Unknown}) and changed the types if needed. This task was very time consuming as the mean number of objects detected amounted to 4000 per single $55' \times 55'$ field (up to 14\,000 at low galactic latitude). This visual quality control check was necessary to reject cases of blended stellar images, which can occur at $m_B < 17.5$ with a non-negligible frequency, especially at increasingly lower galactic latitudes (b $<$ 45\degr).

Finally, we also used POSS-II red (IIIa-F) plates of all our {\sc Galaxy} objects to perform a second check of our final catalogues of neighbours (55\,154 stamps, visually checked by L. V.-M.). The choice of POSS-II instead of POSS-I for this final check removed the detected plate defects in the POSS-I survey that could have passed through our first revision and provided a better spatial resolution to distinguish compact galaxies from stars. We summarise the results of this second visual inspection of the {\sc Galaxy} objects: 98\% were confirmed as {\sc Galaxy} ($\sim$ 54\,000 objects), almost 2\% were plate defects (1119 sources), while 0.04\% were {\sc Star} (23 objects).

\section{Redshifts of the catalogued neighbours} \label{sec:red}

In order to evaluate the physical association of the (projected) neighbours with the CIG galaxies, we searched for available redshifts in the bibliography. As explained above, \citeauthor{1973AISAO...8....3K} did not use redshifts to produce the CIG, since at that time few of such data existed. Nowadays we are able to use more than a dozen databases and surveys in order to search for the redshifts and determine whether the catalogued neighbours are physically associated with the CIG galaxies or just projected objects. We used batch routines for all the 54\,000 neighbours to access each database, matching the coordinates within a tolerance of 6\arcs.

\subsection{Redshift compilation} \label{subsec:redComp}

We compiled all the data coming from the various databases (listed in the first column of Table~\ref{tab:types2}). We treated the different formats in order to obtain one single, homogeneous (J2000 coordinates, heliocentric velocities) final catalogue. A total of 16\,126 (29.9\%) objects have a redshift listed in at least one database. The second column of Table~\ref{tab:types2} lists, for each database, the total number of available redshifts.

The typical reported error on the velocities is about $\sim$ 40\kms. For some galaxies, the redshifts were listed several times, in various databases. The agreement is generally very good (less than tens of \kms) between the different databases. Only one redshift per neighbour was kept for the following study. To have the most homogeneous final database, we chose to preferentially keep the data from the largest surveys. The Sloan Digital Sky Survey (SDSS) gave 12166 objects (75\% of the redshift sample) and besides this it gave the smallest error and the most confident data. Next, in order, we used: the 2dF, the CfA (velocity), NED, HyperLEDA, UZC. Because of the redundancy, the UZCJ2000, CfA1, NOG4 and SSRS2 were not used.

\subsection{Classification of neighbours} \label{subsec:ClassNeig}

This search not only provides the redshift but also in some cases a classification as star/galaxy, providing a third check of our results (see Sect.~\ref{sssec:checks} for the first and second checks). NED, HyperLEDA, SDSS, CfA give types for the objects in their databases, and in some cases, even when no redshift is available. The statistical significance of the type is improved as some objects without redshift have a determined type: the third column of Table~\ref{tab:types2} shows the number of objects with a known type, and the fourth column gives the percentage of {\sc Galaxy} with respect to this number. The CIG neighbours are classified as {\sc Galaxy} in more than 99.90\% of the cases. These results validate our method to separate galaxies from stars.

\begin{table}
\begin{center}
\caption{Databases and surveys searched for the redshifts of the neighbour galaxies (see Sect.~\ref{subsec:redComp} and Sect.~\ref{subsec:ClassNeig}).} \label{tab:types2}
\begin{tabular}{l r r c}
\hline \hline
(1)             & (2)   & (3)     &     (4)\\
Database & Number of &	Number of & Percentage of\\
or survey & redshifts& matched objects & {\sc Galaxy}\\
\hline
NED		& 8024	& 35317	& 99.97\%	\\
hyperLEDA 	& 11608	& 25614	& 99.99\%	\\
SDSS-DR3	& 12166	& 12166	& 99.79\%	\\
CfA (velocity)	& 8864	& 9103	& 99.86\%	\\
2dF		& 3018	& 3018	& - 		\\
UZC		& 1461	& 1488	& -		\\
UZC J2000	& 1445	& 1485	& -		\\
CfA2		& 866	& 866	& 100\%		\\
CfA1		& 106	& 106	& 100\%		\\
NOG2		& 67	& 67	& -		\\
NOG4		& 66	& 66	& -		\\
SSRS2		& 50	& 50	& -		\\
\hline
\end{tabular}
\end{center}
\end{table}

\subsection{Catalogue of neighbour galaxies}

The parameters kept for each {\sc Galaxy} were stored in the form of ASCII catalogues; as an example, the first lines of the catalogues associated with CIG 1 and CIG 2 are shown in Table~\ref{tab:comp}. The entries are:
\begin{itemize}
\item {\it Column 1}: CIG number;
\item {\it Column 2}: Neighbour number;
\item {\it Column 3}: Right Ascension (Epoch J2000);
\item {\it Column 4}: Declination (Epoch J2000);
\item {\it Column 5}: Logarithm of the area of the galaxy (arcsec.$^2$);
\item {\it Column 6}: Apparent magnitude given by SExtractor (MAG\_ISO parameter);
\item {\it Column 7}: Projected distance in \arcs\ between the neighbour and the associated CIG galaxy;
\item {\it Column 8}: Diameter ($D_{25}$) of the neighbour galaxy in \arcs;
\item {\it Columns 9} \& {\it 10}: \citeauthor{1973AISAO...8....3K}'s criterion flags, see Sect.~\ref{sub:kara01};
\item {\it Column 11}: Recession velocity (in \kms) of the neighbour galaxy, when available;
\item {\it Column 12}: Reference for the recession velocity: ``1'' refers to SDSS, ``2'' to 2dF, ``3'' to Velo, ``4'' to NED, ``5'' to HyperLEDA, ``6'' to UZC, ``7'' to CfA2 and ``8'' to NOG2.
\end{itemize}

The parameters used during the SExtraction make the diameters of the detected objects about two times smaller than the expected estimation of $D_{25}$, since the typical 3-sigma for a POSS-I Schmidt plate background results in a brighter detection level of $\mu_b \approx 23.5$ mags/sq. arcsec. In each field, we calculated the scale factor between the known $D_{25}$ (from NED) and the SExtracted value of the CIG diameter. We applied this scale factor to the diameters of the neighbours in order to have an estimated value of their true $D_{25}$. When the scale factor was outside 2 $\sigma$ from the mean factor calculated with the CIG galaxies, we decided to replace it by the mean value (equal to 2). Hence, in these fields, the SExtracted factors of the neighbours were multiplied by 2 to infer the values of the $D_{25}$.

\begin{figure}
\includegraphics[width=\columnwidth]{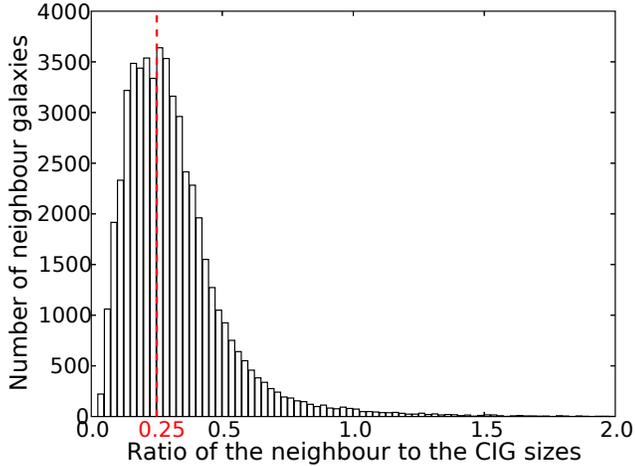}
\caption{Distribution of the major axis of the neighbours with respect to the major axis of their associated CIG galaxy. The red dashed line represents the lower limit defined by \citeauthor{1973AISAO...8....3K} to consider a neighbour as a potential perturber.} \label{fig:histSize}
\end{figure}

In Fig.~\ref{fig:histSize}, we show the distribution of the size of the neighbour galaxies with respect to the size of the associated CIG galaxy. Very few neighbours have a diameter larger than the diameter of their associated CIG galaxy. The distribution increases exponentially as the diameters of the neighbours get smaller: the peak of the distribution is reached for the neighbours having diameters of about one fourth the size of the diameter of the associated CIG galaxy. This corresponds to the nominal factor used by \citeauthor{1973AISAO...8....3K}: the sizes of the neighbours taken into account by \citeauthor{1973AISAO...8....3K}'s isolation criterion are not equally distributed between 0.25 and 4$D_p$, the vast majority of the neighbours (about 88\%) are at least two times smaller than their associated CIG galaxy.

Taking also into account the neighbour galaxies having a diameter less than 0.25 $D_p$ allows us to go a step further and not only exclude major interactions but to establish a gradient in the degree of isolation with respect to small satellites. This will be the subject of a further article \citep[in prep.]{2007A&A...2..2V}.

\begin{table*}
\caption{Catalogue of the neighbours of the CIG galaxies$^\dag$.} \label{tab:comp}
\begin{tabular}{c c c c c c c c c c c c}
\hline \hline
(1) &  (2)   & (3)& (4)  &   (5)     &    (6)   &    (7)   &    (8)   &     (9)     &   (10)   &    (11)  & (12)\\
CIG & Neigh. & RA (\degr) & Dec. (\degr) & Log(area) & MAG\_ISO & Distance & $D_{25}$ & $<$20$D_i$ & factor 4 & Velocity & Ref.\\
    &       & (J2000) & (J2000) & (arcsec.$^2$) & (mag) & (\arcs) & (\arcs) &             &          &  (\kms)  &     \\
\hline
1 & 0 & 0.341470 & -2.361358 & 2.231 & 15.626 & 2237.1 & 22.6 & 0 & 0 & 0 & 0 \\
1 & 1 & 0.393433 & -2.356566 & 2.029 & 17.378 & 2098.6 & 16.4 & 0 & 0 & 0 & 0 \\
1 & 2 & 1.034508 & -2.356130 & 1.979 & 17.380 & 1849.5 & 16.1 & 0 & 0 & 0 & 0 \\
1 & 3 & 1.024575 & -2.346702 & 2.063 & 17.123 & 1802.2 & 17.5 & 0 & 0 & 0 & 0 \\
1 & 4 & 0.781704 & -2.312119 & 2.159 & 17.030 & 1434.1 & 19.3 & 0 & 0 & 0 & 0 \\
1 & 5 & 1.174220 & -2.279433 & 2.529 & 16.028 & 1953.6 & 32.3 & 0 & 1 & 0 & 0 \\
\vdots & \vdots & \vdots & \vdots & \vdots & \vdots & \vdots & \vdots & \vdots & \vdots & \vdots & \vdots\\
2 & 0 & 1.112587 & 29.343813 & 2.287 & 16.384 & 1850.9 & 17.4 & 0 & 1 & 0 & 0 \\
2 & 1 & 0.627120 & 29.414143 & 2.231 & 16.420 & 1525.4 & 16.1 & 0 & 1 & 0 & 0 \\
2 & 2 & 0.467379 & 29.498917 & 2.123 & 16.904 & 1573.5 & 15.0 & 0 & 1 & 0 & 0 \\
2 & 3 & 0.467854 & 29.505060 & 1.966 & 16.957 & 1557.4 & 15.3 & 0 & 1 & 0 & 0 \\
2 & 4 & 1.149291 & 29.516792 & 2.287 & 16.435 & 1410.9 & 18.4 & 0 & 1 & 0 & 0 \\
2 & 5 & 0.735791 & 29.584400 & 2.347 & 16.087 &  827.2 & 24.5 & 0 & 1 & 4836 & 3 \\
\vdots & \vdots & \vdots & \vdots & \vdots & \vdots & \vdots & \vdots & \vdots & \vdots & \vdots & \vdots\\
\hline
\end{tabular}

$^\dag$The full table is available in electronic form at CDS or from {\tt http://www.iaa.csic.es/AMIGA.html}.
\end{table*}

\section{Discussion} \label{sec:disc}

\subsection{Karachentseva's criterion in light of available new information} \label{sub:kara01}

For practical reasons we could not cover fields as large as the needed ones to fully verify \citeauthor{1973AISAO...8....3K}'s criterion, but we were still able to find some of the CIG galaxies that failed her criterion. According to \citeauthor{1973AISAO...8....3K}, a perturbative neighbour can be 4 times bigger and $20 D_i$ away from the CIG galaxy. This is a huge distance: $20 D_i = 20 \times 4 D_p = 80 D_p$. We could only cover this area for 74 fields: among them, 58 CIG galaxies are isolated following \citeauthor{1973AISAO...8....3K}'s criterion, while 16 are not isolated.

For the remaining fields, although we were not able to check the whole $80 \times D_p$, we have found 284 CIG galaxies violating \citeauthor{1973AISAO...8....3K}'s isolation definition. Still, 666 CIG galaxies remain isolated accordingly to \citeauthor{1973AISAO...8....3K}, taking into account that we cannot assert that some of these latter galaxies will not move from the ``isolated'' to the ``not isolated'' sample, if studying a larger field.

The majority of the neighbours (30\,407 galaxies, 57.3\%) have sizes similar (within a factor of 4) to the one of their associated CIG galaxy, but only 1.4\% of companions (734) are within 20 times of their diameters away from the CIG galaxy. We find that 465 neighbours cumulate the two conditions, hence violating \citeauthor{1973AISAO...8....3K}'s criterion. As several of these neighbours could be in the same field around one given CIG galaxy, a total of 284 CIG galaxies were concerned.

Two columns of Table~\ref{tab:comp} summarise these conditions: {\it Column 9} is ``1'' if the neighbour is within 20 $D_i$ and equal to ``0'' if it is farther away; {\it Column 10} is ``1'' if the neighbour has a diameter similar to the one of the associated CIG galaxy (factor 4 in size) and equal to ``0'' otherwise. If at least one of the two conditions is false, multiplying the last two columns, we find 0 which means that the CIG galaxy is still isolated according to \citeauthor{1973AISAO...8....3K}'s criterion; if the two conditions are true, the multiplication gives 1: the CIG galaxy it is not isolated according to \citeauthor{1973AISAO...8....3K}'s criterion.

\subsection{Redshifts}

\begin{figure}
\includegraphics[width=\columnwidth]{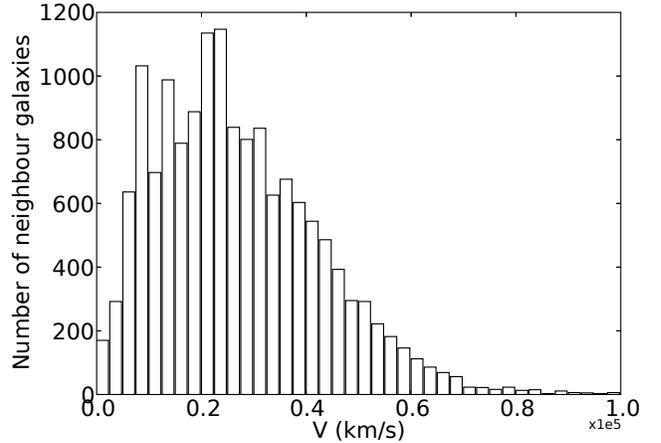}
\caption{Distribution of the available recession velocities for the neighbour galaxies.} \label{fig:reds}
\end{figure}

The distribution of the velocities available for the neighbour galaxies is presented in Fig.~\ref{fig:reds}. The mean recession velocity for the neighbour galaxies is about 27\,000\kms. Comparing this value with the distribution of the CIG galaxies' recession velocities (Fig.~\ref{fig:histVelCig}) showing a mean at about 6624\kms, it appears that the neighbour galaxies represent a deeper sample of galaxies than the CIG. Hence, most of the neighbour galaxies are background galaxies and although some of the CIG galaxies violate \citeauthor{1973AISAO...8....3K}'s strict criterion, most of them still represent a valuable population of isolated galaxies.

\subsection{Pair candidates}

As almost one third of the CIG galaxies failed \citeauthor{1973AISAO...8....3K}'s original criterion, we chose to lead a systematic study to identify CIG galaxies possibly belonging to a pair system. The pair candidates are defined as a CIG galaxy with at least one neighbour (factor 2 in size with respect to $D_p$) within $5 \times D_p$. Table~\ref{tab:pair} lists the 10 pair candidates found accordingly. The entries are:
\begin{itemize}
\item {\it Column 1}: CIG number and neighbour number;
\item {\it Column 2}: Right Ascension (in \degr, Epoch J2000);
\item {\it Column 3}: Declination (in \degr, Epoch J2000);
\item {\it Column 4}: Projected distance (in \arcs) from the neighbour to its associated CIG galaxy;
\item {\it Column 5}: Diameter of the galaxy (in \arcs);
\item {\it Column 6}: Recession velocity (when available, in \kms) of the galaxy.
\end{itemize}

CIG 19 has 2 neighbours nearby, one without known velocity and one with a recession velocity very similar to the one of the CIG galaxy: this constitutes a physical pair. Among the 4 other pair candidates having velocity information, 3 CIG galaxies are physically associated with their neighbours (CIGs 74, 488, 533) while this is clearly not the case for CIG 683 (velocity difference of $\sim$ 10\,000 km s$^{-1}$).

Unfortunately, no velocities are available for the neighbours of the 5 remaining pair candidates (CIGs 36, 178, 233, 315, 934). But, as four out of five pair candidates appeared to be real pairs when the velocity is known, we can expect that, again, about 80\% of the 5 pair candidates would be physically bounded.

\begin{table}
{\scriptsize
\caption{Pair candidates.} \label{tab:pair}
\begin{tabular}{c c c c c r}
\hline \hline
   (1)   &     (2)    &    (3)    &    (4)   &   (5)    &   (6)   \\
Galaxy   &     RA     &   Dec.    & Distance & Diameter & Velocity\\
         & (\degr)  & (\degr) & (\arcs) & (\arcs) &(km s$^{-1}$)\\
\hline
{\bf CIG 19} &   6.067841 & 14.237000    &    & 54.0 & 5390\\
Neigh. 20 &   6.074004 & 14.272449 & 129.3 & 32.7 & 5396\\
Neigh. 22 &   6.130088 & 14.260384 & 234.6 & 39.1 & No data\\
\hline
{\bf CIG 36} &  12.861758 & 40.725868 &    & 60.0 & 5855\\
Neigh.  8 &  12.952467 & 40.762981 & 282.3 & 36.5 & No data\\
\hline
{\bf CIG 74} &  29.330297 & 28.590328 &    & 36.0 & 10188\\
Neigh. 62 &  29.314213 & 28.614264 & 100.0 & 29.1 & 10300\\
\hline
{\bf CIG 178} & 107.163582 & 61.305061 &    & 18.0 & 7610\\
Neigh. 17 & 107.11628  & 61.299938 &  82.8 & 10.9 & No data\\
\hline
{\bf CIG 233} & 122.907974 & 27.538559 &    & 24.0 & 11225\\
Neigh. 21 & 122.879021 & 27.524349 & 104.2 & 12.1 & No data\\
\hline
{\bf CIG 315} & 137.892471 & -3.536764 &    & 54.0 & 5088\\
Neigh. 26 & 137.853882 & -3.599669 & 265.2 & 33.3 & No data\\
\hline
{\bf CIG 488} & 173.924164 & 73.452034 &    & 84.0 & 12501\\
Neigh. 35 & 174.137344 & 73.470009 & 229.6 & 56.9 & 12425\\
\hline
{\bf CIG 533} & 187.935638 & -1.010247 &    & 24.0 & 21663\\
Neigh. 93 & 187.933319 & -1.005513 &  18.8 & 13.5 & 21585\\
\hline
{\bf CIG 683} & 232.688354 & -0.369905 &    & 36.0 & 11362\\
Neigh. 53 & 232.679489 & -0.383188 &  57.1 & 20.0 & 21285\\
\hline
{\bf CIG 934} & 328.329865 & -2.225402 &    & 42.0 & 5378\\
Neigh. 33 & 328.308563 & -2.192905 & 138.8 & 25.0 & No data\\
\hline
\end{tabular}
}
\end{table}

\subsection{Difference in velocity for a factor 4 in size}

\begin{figure}
\includegraphics[width=\columnwidth]{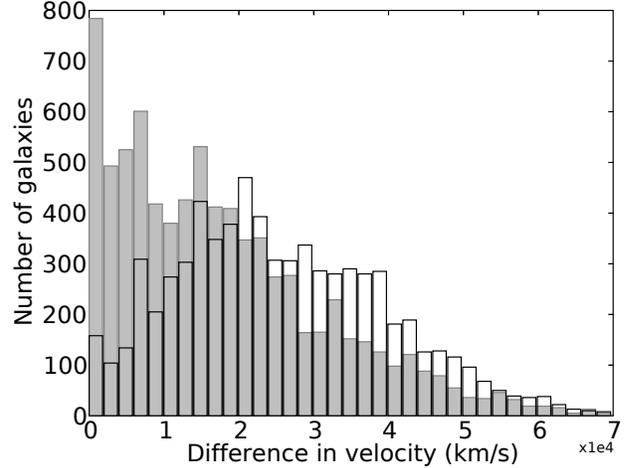}
\caption{Comparison of the velocity difference distributions for the neighbour galaxies considered by \citeauthor{1973AISAO...8....3K}'s criterion (factor 4 in size with respect to their associated CIG galaxy, grey histogram) and for the remaining neighbours (outside the factor 4 in size, unfilled histogram).} \label{fig:velocities}
\end{figure}

In order to determine some of the characteristics of the neighbour population considered by the \citeauthor{1973AISAO...8....3K}'s criterion, we have been able to estimate to what velocity difference the factor 4 in size defining the original isolation criterion corresponds. First, we can see (Fig.~\ref{fig:velocities}) that the similar size (factor 4 in size with respect to the associated CIG galaxy) population dominates up to a velocity difference of 20\,000\kms\ (the mean value is about 17\,700\kms). This high value for the velocity difference between the neighbours and the CIG galaxies makes the CIG a very restrictive sample compared to others (mainly cuts at 500 or 1000\kms). For velocity differences larger than 20\,000\kms, the other sample (neighbours having a size smaller than 0.25$D_p$ or greater than 4$D_p$) dominates: the mean value is about 25\,700\kms.

Second, the distributions of the two samples appear mixed due to the intrinsic recession velocity of each galaxy (the standard deviation is quite large: about 14\,000\kms\ for each sample), which is not a bijective function of the size. This indicates that we need to take into account also the small neighbours (size minor than $4 \times D_p$) that could have a velocity similar to the one of their associated CIG galaxy and exert a noticeable influence. A further analysis is needed to take into account this effect: this is discussed in a following paper \citep[in prep.]{2007A&A...2..2V}.



\subsection{Difference in magnitude for a factor 4 in size}

\begin{figure}
\includegraphics[width=\columnwidth]{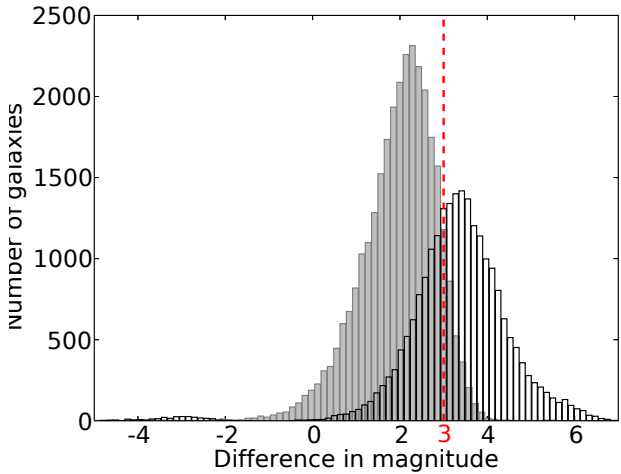}
\caption{Comparison of the magnitude difference distributions for the neighbour galaxies considered by \citeauthor{1973AISAO...8....3K}'s criterion (factor 4 in size with respect to their associated CIG galaxy, grey histogram) and for the remaining neighbours (outside the factor 4 in size, unfilled histogram).} \label{fig:magnitudes}
\end{figure}

The magnitudes for the galaxies catalogued are calibrated on an absolute scale, but it is possible to compare them to each other. Using the magnitude difference between the neighbour galaxies and their associated CIG galaxy removes part of the fluctuation of the zero point from one Schmidt plate to another. \citet{2005AJ....129.2062A} and \citet{2005astro.ph.11432X} claim that an equivalent criterion to \citeauthor{1973AISAO...8....3K}'s one could be obtained by selecting the neighbours within an interval of magnitude equal to 3 with respect to the magnitude of the CIG galaxy, following the equation ($M_{\rm Neig.}$ and $F_{\rm Neig.}$ are the magnitude and flux of a given neighbour; $M_{\rm CIG}$ and $F_{\rm CIG}$ the respective quantities for the CIG galaxy):
\[M_{\rm Neig.} - M_{\rm CIG} = -2.5 \log(\frac{F_{\rm Neig.}}{F_{\rm CIG}})\]
\[\frac{F_{\rm CIG}}{F_{\rm Neig.}} = 10^{3/2.5} = 15.85\]
which is roughly equal to the square of the linear size chosen by \citeauthor{1973AISAO...8....3K} (4$^2$). In Fig.~\ref{fig:magnitudes}, we show the difference in magnitudes with respect to the CIG galaxies for the neighbours that have similar sizes to the CIG galaxies (the mean is 2 and the standard deviation 0.9) and the difference of magnitudes for the rest of the sample (mean is 3.3 and the standard deviation is 1.2). The overlap between the two distributions shows that a cut in magnitude at 3 is a rather good approximation because it loses only 10\% of the neighbours selected by \citeauthor{1973AISAO...8....3K} on the basis of the linear size of 4. But the contamination also shows that the hypothesis of flat surface brightness profile of galaxies is not always true: the cut in magnitude includes a large number of galaxies not considered by \citeauthor{1973AISAO...8....3K}. Hence, the two definitions to seek for the neighbours are not fully equivalent.






\section{Summary and conclusions} \label{sec:conc}

We have performed a systematic study to list all the projected neighbours, down to a magnitude  $m \sim 17.5$, lying within 0.5~Mpc around the CIG galaxies. Our conclusions are the following:

\begin{enumerate}
\item Using automated classification, some of the galaxies in the CIG appear not to be isolated according to \citeauthor{1973AISAO...8....3K}'s original criterion. Nevertheless, this remains a valid sample as a starting point to obtain a refined sample of isolated galaxies. We give a first catalogue distinguishing the isolated CIG galaxies according to \citeauthor{1973AISAO...8....3K} from those failing the original criterion: a systematic and visual inspection of the objects lead to about 54\,000 neighbours.
\item We looked for the available redshifts of the neighbour galaxies in 12 databases: 30\% have known velocities, and the neighbour galaxies represent statistically a background population, with respect to the CIG galaxies.
\item We identified some physical pairs and pair candidates. When a neighbour galaxy is found very near in projection from the central CIG galaxy, its probability of being a physical companion is high.
\item The factor 4 in size defined in the original criterion takes into account the majority of galaxies up to a velocity difference of 20\,000\kms, which makes the CIG original criterion much more restrictive than commonly used criteria which use velocity differences of about 1000\kms.
\item Criteria based on magnitude differences to select the neighbours are not fully consistent with \citeauthor{1973AISAO...8....3K}'s criterion based on the linear sizes. The discrepancy arises from the fact that the surface brightness profile of galaxies is not flat, but nevertheless they are similar criteria to a first approximation.
\item A catalogue of the $\sim$ 54\,000 neighbour galaxies is available electronically at the CDS with positions, magnitudes, areas as well as redshifts when available.
\end{enumerate}

\begin{acknowledgements}
S. V. thanks Gary A. Mamon, Chantal Balkowski, Alessandro Boselli, Santiago Garcia-Burillo, Jos\'e M. Vilchez for their comments and J.-A. Jim\'enez Madrid for the Debian support. This research has made use of the NASA/IPAC Extragalactic Database (NED) which is operated by the Jet Propulsion Laboratory, California Institute of Technology, under contract with the National Aeronautics and Space Administration, and of the Lyon Extragalactic Database (LEDA). This work has been partially supported by DGI Grant AYA 2005-07516-C02-01 and the Junta de Andaluc\'{\i}a (Spain). UL acknowledges support by the research project ESP\,2004-06870-C02-02. Jack Sulentic is partially supported by a sabbatical grant SAB2004-01-04 of the Spanish Ministerio de Educaci\'on y Ciencias. GB acknowledges support at the IAA/CSIC by an I3P contract (I3P-PC2005F) funded by the European Social Fund.
\end{acknowledgements}

\bibliography{ref}

\begin{thebibliography}{61}
\expandafter\ifx\csname natexlab\endcsname\relax\def\natexlab#1{#1}\fi

\bibitem[{{Aars} {et~al.}(2001){Aars}, {Marcum}, \&
  {Fanelli}}]{2001AJ....122.2923A}
{Aars}, C.~E., {Marcum}, P.~M., \& {Fanelli}, M.~N. 2001, \aj, 122, 2923

\bibitem[{{Adams} {et~al.}(1980){Adams}, {Jensen}, \&
  {Stocke}}]{1980AJ.....85.1010A}
{Adams}, M.~T., {Jensen}, E.~B., \& {Stocke}, J.~T. 1980, \aj, 85, 1010

\bibitem[{{Aguerri}(1999)}]{1999A&A...351...43A}
{Aguerri}, J.~A.~L. 1999, \aap, 351, 43

\bibitem[{{Allam} {et~al.}(2005){Allam}, {Tucker}, {Lee}, \&
  {Smith}}]{2005AJ....129.2062A}
{Allam}, S.~S., {Tucker}, D.~L., {Lee}, B.~C., \& {Smith}, J.~A. 2005, \aj,
  129, 2062

\bibitem[{{Allen} {et~al.}(1973){Allen}, {Ekers}, {Burke}, \&
  {Miley}}]{1973Natur.241..260A}
{Allen}, R.~J., {Ekers}, R.~D., {Burke}, B.~F., \& {Miley}, G.~K. 1973, \nat,
  241, 260

\bibitem[{{Balkowski} \& {Chamaraux}(1981)}]{1981A&A....97..223B}
{Balkowski}, C. \& {Chamaraux}, P. 1981, \aap, 97, 223

\bibitem[{{Bertin} \& {Arnouts}(1996)}]{1996A&AS..117..393B}
{Bertin}, E. \& {Arnouts}, S. 1996, \aaps, 117, 393

\bibitem[{{Colbert} {et~al.}(2001){Colbert}, {Mulchaey}, \&
  {Zabludoff}}]{2001AJ....121..808C}
{Colbert}, J.~W., {Mulchaey}, J.~S., \& {Zabludoff}, A.~I. 2001, \aj, 121, 808

\bibitem[{{Espada}(2006)}]{2006.PhD.Thesis.E}
{Espada}, D. 2006, PhD thesis, Instituto de Astrof\'isica de Andaluc\'ia

\bibitem[{{Espada} {et~al.}(2005){Espada}, {Bosma}, {Verdes-Montenegro},
  {Athanassoula}, {Leon}, {Sulentic}, \& {Yun}}]{2005A&A...442..455E}
{Espada}, D., {Bosma}, A., {Verdes-Montenegro}, L., {et~al.} 2005, \aap, 442,
  455

\bibitem[{{Grebel}(2006)}]{2006astro.ph..5564G}
{Grebel}, E.~K. 2006, ArXiv Astrophysics e-prints

\bibitem[{{Haynes} \& {Giovanelli}(1983)}]{1983ApJ...275..472H}
{Haynes}, M.~P. \& {Giovanelli}, R. 1983, \apj, 275, 472

\bibitem[{{Haynes} \& {Giovanelli}(1984)}]{1984AJ.....89..758H}
{Haynes}, M.~P. \& {Giovanelli}, R. 1984, \aj, 89, 758

\bibitem[{{Huchra} \& {Thuan}(1977)}]{1977ApJ...216..694H}
{Huchra}, J. \& {Thuan}, T.~X. 1977, \apj, 216, 694

\bibitem[{{Karachentsev} {et~al.}(1996){Karachentsev}, {Musella}, \&
  {Grimaldi}}]{1996A&A...310..722K}
{Karachentsev}, I., {Musella}, I., \& {Grimaldi}, A. 1996, \aap, 310, 722

\bibitem[{{Karachentsev}(1972)}]{1972SoSAO...7....1K}
{Karachentsev}, I.~D. 1972, Soobshcheniya Spetsial'noj Astrofizicheskoj
  Observatorii, 7, 1

\bibitem[{{Karachentseva}(1973)}]{1973AISAO...8....3K}
{Karachentseva}, V.~E. 1973, Astrofizicheskie Issledovaniia Izvestiya
  Spetsial'noj Astrofizicheskoj Observatorii, 8, 3

\bibitem[{{Karachentseva}(1980)}]{1980SvA....24..665K}
{Karachentseva}, V.~E. 1980, Soviet Astronomy, 24, 665

\bibitem[{{Kornreich} {et~al.}(2001){Kornreich}, {Haynes}, {Jore}, \&
  {Lovelace}}]{2001AJ....121.1358K}
{Kornreich}, D.~A., {Haynes}, M.~P., {Jore}, K.~P., \& {Lovelace}, R.~V.~E.
  2001, \aj, 121, 1358

\bibitem[{{Larson} \& {Tinsley}(1978)}]{1978ApJ...219...46L}
{Larson}, R.~B. \& {Tinsley}, B.~M. 1978, \apj, 219, 46

\bibitem[{{Leon} {et~al.}(2000){Leon}, {Meylan}, \&
  {Combes}}]{2000A&A...359..907L}
{Leon}, S., {Meylan}, G., \& {Combes}, F. 2000, \aap, 359, 907

\bibitem[{{Leon} \& {Verdes-Montenegro}(2003)}]{2003A&A...411..391L}
{Leon}, S. \& {Verdes-Montenegro}, L. 2003, \aap, 411, 391

\bibitem[{{Lisenfeld} {et~al.}(2007){Lisenfeld}, {Verdes-Montenegro},
  {Sulentic}, {Leon}, {Espada}, {Bergond}, {Garc{\'{\i}}a}, {Sabater},
  {Santander-Vela}, \& {Verley}}]{2007A&A...462..507L}
{Lisenfeld}, U., {Verdes-Montenegro}, L., {Sulentic}, J., {et~al.} 2007, \aap,
  462, 507

\bibitem[{{M{\' a}rquez} {et~al.}(2002){M{\' a}rquez}, {Masegosa}, {Moles},
  {Varela}, {Bettoni}, \& {Galletta}}]{2002A&A...393..389M}
{M{\' a}rquez}, I., {Masegosa}, J., {Moles}, M., {et~al.} 2002, \aap, 393, 389

\bibitem[{{M{\' a}rquez} {et~al.}(2003){M{\' a}rquez}, {Masegosa}, {Moles},
  {Varela}, {Bettoni}, \& {Galletta}}]{2003Ap&SS.284..711M}
{M{\' a}rquez}, I., {Masegosa}, J., {Moles}, M., {et~al.} 2003, \apss, 284, 711

\bibitem[{{M{\' a}rquez} \& {Moles}(1996)}]{1996A&AS..120....1M}
{M{\' a}rquez}, I. \& {Moles}, M. 1996, \aaps, 120, 1

\bibitem[{{M{\' a}rquez} \& {Moles}(1999)}]{1999A&A...344..421M}
{M{\' a}rquez}, I. \& {Moles}, M. 1999, \aap, 344, 421

\bibitem[{{Madore} {et~al.}(2004){Madore}, {Freedman}, \&
  {Bothun}}]{2004ApJ...607..810M}
{Madore}, B.~F., {Freedman}, W.~L., \& {Bothun}, G.~D. 2004, \apj, 607, 810

\bibitem[{{Marcum} {et~al.}(2004){Marcum}, {Aars}, \&
  {Fanelli}}]{2004AJ....127.3213M}
{Marcum}, P.~M., {Aars}, C.~E., \& {Fanelli}, M.~N. 2004, \aj, 127, 3213

\bibitem[{{Morgan} {et~al.}(1998){Morgan}, {Smith}, \&
  {Phillipps}}]{1998MNRAS.295...99M}
{Morgan}, I., {Smith}, R.~M., \& {Phillipps}, S. 1998, \mnras, 295, 99

\bibitem[{{Odewahn}(1995)}]{1995PASP..107..770O}
{Odewahn}, S.~C. 1995, \pasp, 107, 770

\bibitem[{{Odewahn} {et~al.}(2002){Odewahn}, {Cohen}, {Windhorst}, \&
  {Philip}}]{2002ApJ...568..539O}
{Odewahn}, S.~C., {Cohen}, S.~H., {Windhorst}, R.~A., \& {Philip}, N.~S. 2002,
  \apj, 568, 539

\bibitem[{{Odewahn} {et~al.}(1996){Odewahn}, {Windhorst}, {Driver}, \&
  {Keel}}]{1996ApJ...472L..13O}
{Odewahn}, S.~C., {Windhorst}, R.~A., {Driver}, S.~P., \& {Keel}, W.~C. 1996,
  \apjl, 472, L13+

\bibitem[{{Peebles}(1974{\natexlab{a}})}]{1974ApJ...189L..51P}
{Peebles}, P.~J.~E. 1974{\natexlab{a}}, \apjl, 189, L51+

\bibitem[{{Peebles}(1974{\natexlab{b}})}]{1974A&A....32..197P}
{Peebles}, P.~J.~E. 1974{\natexlab{b}}, \aap, 32, 197

\bibitem[{{Pisano} \& {Wilcots}(1999)}]{1999AJ....117.2168P}
{Pisano}, D.~J. \& {Wilcots}, E.~M. 1999, \aj, 117, 2168

\bibitem[{{Pisano} {et~al.}(2002){Pisano}, {Wilcots}, \&
  {Liu}}]{2002ApJS..142..161P}
{Pisano}, D.~J., {Wilcots}, E.~M., \& {Liu}, C.~T. 2002, \apjs, 142, 161

\bibitem[{{Prada} {et~al.}(2003){Prada}, {Vitvitska}, {Klypin}, {Holtzman},
  {Schlegel}, {Grebel}, {Rix}, {Brinkmann}, {McKay}, \&
  {Csabai}}]{2003ApJ...598..260P}
{Prada}, F., {Vitvitska}, M., {Klypin}, A., {et~al.} 2003, \apj, 598, 260

\bibitem[{{Reda} {et~al.}(2004){Reda}, {Forbes}, {Beasley}, {O'Sullivan}, \&
  {Goudfrooij}}]{2004MNRAS.354..851R}
{Reda}, F.~M., {Forbes}, D.~A., {Beasley}, M.~A., {O'Sullivan}, E.~J., \&
  {Goudfrooij}, P. 2004, \mnras, 354, 851

\bibitem[{{Schmidt}(1968)}]{1968ApJ...151..393S}
{Schmidt}, M. 1968, \apj, 151, 393

\bibitem[{{Soneira} \& {Peebles}(1977)}]{1977ApJ...211....1S}
{Soneira}, R.~M. \& {Peebles}, P.~J.~E. 1977, \apj, 211, 1

\bibitem[{{Stocke}(1978)}]{1978AJ.....83..348S}
{Stocke}, J.~T. 1978, \aj, 83, 348

\bibitem[{{Stocke} {et~al.}(1978){Stocke}, {Tifft}, \&
  {Kaftan-Kassim}}]{1978AJ.....83..322S}
{Stocke}, J.~T., {Tifft}, W.~G., \& {Kaftan-Kassim}, M.~A. 1978, \aj, 83, 322

\bibitem[{{Sulentic}(1976)}]{1976ApJS...32..171S}
{Sulentic}, J.~W. 1976, \apjs, 32, 171

\bibitem[{{Sulentic} {et~al.}(2006){Sulentic}, {Verdes-Montenegro}, {Bergond},
  {Lisenfeld}, {Durbala}, {Espada}, {Garcia}, {Leon}, {Sabater}, {Verley},
  {Casanova}, \& {Sota}}]{2006A&A...449..937S}
{Sulentic}, J.~W., {Verdes-Montenegro}, L., {Bergond}, G., {et~al.} 2006, \aap,
  449, 937

\bibitem[{{Toledo} {et~al.}(1999){Toledo}, {Dultzin-Hacyan}, {Gonzalez}, \&
  {Sulentic}}]{1999AJ....118..108T}
{Toledo}, H.~M.~H., {Dultzin-Hacyan}, D., {Gonzalez}, J.~J., \& {Sulentic},
  J.~W. 1999, \aj, 118, 108

\bibitem[{{Toomre}(1977)}]{1977ARA&A..15..437T}
{Toomre}, A. 1977, \araa, 15, 437

\bibitem[{{Toomre} \& {Toomre}(1972)}]{1972ApJ...178..623T}
{Toomre}, A. \& {Toomre}, J. 1972, \apj, 178, 623

\bibitem[{{Turner} \& {Gott}(1975)}]{1975ApJ...197L..89T}
{Turner}, E.~L. \& {Gott}, J.~R. 1975, \apjl, 197, L89

\bibitem[{{Varela} {et~al.}(2004){Varela}, {Moles}, {M{\' a}rquez}, {Galletta},
  {Masegosa}, \& {Bettoni}}]{2004A&A...420..873V}
{Varela}, J., {Moles}, M., {M{\' a}rquez}, I., {et~al.} 2004, \aap, 420, 873

\bibitem[{{Verdes-Montenegro} {et~al.}(1995){Verdes-Montenegro}, {Bosma}, \&
  {Athanassoula}}]{1995A&A...300...65V}
{Verdes-Montenegro}, L., {Bosma}, A., \& {Athanassoula}, E. 1995, \aap, 300, 65

\bibitem[{{Verdes-Montenegro} {et~al.}(1997){Verdes-Montenegro}, {Bosma}, \&
  {Athanassoula}}]{1997A&A...321..754V}
{Verdes-Montenegro}, L., {Bosma}, A., \& {Athanassoula}, E. 1997, \aap, 321,
  754

\bibitem[{{Verdes-Montenegro} {et~al.}(2005){Verdes-Montenegro}, {Sulentic},
  {Lisenfeld}, {Leon}, {Espada}, {Garcia}, {Sabater}, \&
  {Verley}}]{2005A&A...436..443V}
{Verdes-Montenegro}, L., {Sulentic}, J., {Lisenfeld}, U., {et~al.} 2005, \aap,
  436, 443

\bibitem[{{Verley}(2005)}]{2005.PhD.Thesis.V}
{Verley}, S. 2005, PhD thesis, Observatoire de Paris - Instituto de
  Astrof\'isica de Andaluc\'ia

\bibitem[{{Verley} {et~al.}(2007){Verley}, {Leon}, {Verdes-Montenegro},
  {Combes}, {Sabater}, {Sulentic}, {Bergond}, {Espada}, {Garc\'ia},
  {Lisenfeld}, \& {Odewahn}}]{2007A&A...2..2V}
{Verley}, S., {Leon}, S., {Verdes-Montenegro}, L., {et~al.} 2007, \aap,
  submitted

\bibitem[{{Vettolani} {et~al.}(1986){Vettolani}, {de Souza}, \&
  {Chincarini}}]{1986A&A...154..343V}
{Vettolani}, G., {de Souza}, R., \& {Chincarini}, G. 1986, \aap, 154, 343

\bibitem[{{Xanthopoulos} \& {de Robertis}(1991)}]{1991AJ....102.1980X}
{Xanthopoulos}, E. \& {de Robertis}, M.~M. 1991, \aj, 102, 1980

\bibitem[{{Xinfa} {et~al.}(2005){Xinfa}, {Xinsheng}, {Chenghong}, {Qun}, \&
  {Ji-zhou}}]{2005astro.ph.11432X}
{Xinfa}, D., {Xinsheng}, M., {Chenghong}, L., {Qun}, Z., \& {Ji-zhou}, H. 2005,
  ArXiv Astrophysics e-prints

\bibitem[{{Xu} \& {Sulentic}(1991)}]{1991ApJ...374..407X}
{Xu}, C. \& {Sulentic}, J.~W. 1991, \apj, 374, 407

\bibitem[{{Zaritsky} {et~al.}(1993){Zaritsky}, {Smith}, {Frenk}, \&
  {White}}]{1993ApJ...405..464Z}
{Zaritsky}, D., {Smith}, R., {Frenk}, C., \& {White}, S.~D.~M. 1993, \apj, 405,
  464

\bibitem[{{Zwicky} {et~al.}(1968){Zwicky}, {Herzog}, \&
  {Wild}}]{1968cgcg.book.....Z}
{Zwicky}, F., {Herzog}, E., \& {Wild}, P. 1968, {Catalogue of galaxies and of
  clusters of galaxies} (Pasadena: California Institute of Technology (CIT),
  1961-1968)

\end{thebibliography}

\end{document}